\documentclass[10pt,letterpaper,conference]{ieeeconf}

\usepackage{calrsfs,times}
\usepackage{graphicx,psfrag,enumerate}
\usepackage{amssymb}
\usepackage{amsmath}
\usepackage{calrsfs}
\usepackage{color, epsfig, epstopdf, wrapfig}
\usepackage{paralist}
\usepackage{float,cuted}

\newtheorem{theorem}{Theorem}
\newtheorem{remark}{Remark}

\newtheorem{lemma}{Lemma}

\newtheorem{proposition}{Proposition}

\def\tr{\mathop{\rm tr}\nolimits} 
\usepackage{amsmath}
\usepackage{accents}
\newlength{\dhatheight}

\usepackage{color}
\usepackage[normalem]{ulem}
\usepackage{soul} 
\soulregister\cite7
\soulregister\ref7
\soulregister\pageref7

\IEEEoverridecommandlockouts

\begin{document}

\title{\LARGE \bf Active versus Passive Coherent Equalization of Passive
  Linear Quantum Systems\thanks{This work was 
    supported by the Australian Research Council and the ARC Centre for Quantum
    Computation and Communication Technology.}} 

\author{V.~Ugrinovskii\thanks{V. Ugrinovskii is with the School of
    Engineering and Information Technology, University of New South Wales
    Canberra, Canberra, ACT 2600,
    Australia, {\tt\small v.ougrinovski@adfa.edu.au}} 
  \and 
       M.~R.~James\thanks{M.~R.~James are with the ARC Centre for Quantum
    Computation and Communication Technology, Research School of Engineering,
    the Australian National University, Canberra, ACT 2601, Australia,
  {\tt\small matthew.james@anu.edu.au.}}} 

\maketitle
         
\begin{abstract}
The paper considers the problem of equalization of passive
linear quantum systems. While our previous work was concerned with the
analysis and synthesis of passive equalizers, in this paper we analyze 
coherent quantum equalizers whose annihilation (respectively, creation)
operator dynamics in the Heisenberg picture are driven by both quadratures of 
the channel output field. We show that the characteristics of the input
field must be taken into consideration when choosing the type of the
equalizing filter. In particular, we show that for thermal fields allowing
the filter to process both quadratures of the channel output may not
improve mean square accuracy of the input field estimate, in comparison with passive
filters. This situation changes when the input field is
`squeezed'.  
\end{abstract}

\section{Introduction}

Quantum communication systems are
subject to fundamental quantum mechanical limits which restrict their
capacity to transfer information. Due to these limitations, the 
problem of correcting distortions in quantum communication systems differs
significantly from its classical counterparts. This point has been demonstrated 
in~\cite{UJ2a} where we highlighted some conceptual differences which arise
when Wiener's paradigm of mean-square error
optimization~\cite{Kailath-1981} is  
applied in the derivation of coherent quantum filters
. In particular, 
optimal coherent equalizing filters may require a noise
field to be injected into the filter, and for optimal
performance, the filter must be tuned to balance this noise against the
noise in the channel output. 


In this paper, we continue the analysis of the coherent equalization
problem for quantum communication channels introduced
in~\cite{UJ2a}. Although the
problem resembles the problem of optimal Wiener 
filtering, the coherent equalizer must satisfy 
the laws of quantum physics in that it must preserve certain operator
commutation relations. This leads to additional requirements on the
synthesized equalizing filter, known as 
physical realizability, which do not 
arise in the classical filtering theory~\cite{JNP-2008,SP-2012}. It has
been shown in~\cite{UJ2a} 
that even in the simplified case concerned with equalization of passive
quantum channels using passive coherent filters, these 
additional requirements translate into nontrivial optimization constraints,
and the problem of optimal coherent filtering reduces to a 
challenging nonconvex optimization problem. It has been observed
in~\cite{UJ2a} that an optimal passive
coherent equalizer is not always able to reduce the mean-square
error between the channel input and output fields. This naturally leads to the question as to whether expanding the
class of filters to include more general active filters can help to resolve
this issue. 

As it turns out, the potentially greater flexibility in shaping the filter
output offered by active filters  
is not always easy to realize --- our first
result identifies a class of coherent equalization problems involving
thermal input field in which active
coherent filters have no advantage over passive (noncausal) filters. On the
other hand, our second result demonstrates that when the input field is
`squeezed', 
the mean-square optimal coherent filter utilizes both quadratures of the
channel output.  

The paper is organized as follows. In the next section we present the
necessary basics of linear quantum systems.  The quantum equalization problem is
reviewed in Section~\ref{sec:equal-probl-annih}. 
Next, 
Section~\ref{sec:main-results} presents the results of the paper. 
Concluding remarks 
are presented in Section~\ref{Conclusions}.

\paragraph*{Notation}
For an operator $\mathbf{a}$ in a Hilbert space $\mathfrak{H}$, $\mathbf{a}^*$  denotes the
Hermitian adjoint operator, and if $a$ is a complex number, $a^*$
is its 
complex conjugate. The notation $\mathrm{col}(\mathbf{a},\mathbf{b})$
denotes the column vector of operators obtained by concatenating operators
$\mathbf{a}$ and $\mathbf{b}$. Let
$\mathbf{a}=\mathrm{col}(\mathbf{a}_1,\ldots,\mathbf{a}_n)$ be a 
column vector comprised of $n$ 
operators (i.e., $\mathbf{a}$ is an operator $\mathfrak{H}\to
\mathfrak{H}^n$); then  
$\mathbf{a}^\#=\mathrm{col}(\mathbf{a}_1^*,\ldots,\mathbf{a}_n^*)$, $\mathbf{a}^T=(\mathbf{a}_1^T~\ldots~ \mathbf{a}_n^T)$ (i.e, the row of
operators), and $\mathbf{a}^\dagger = (\mathbf{a}^\#)^T$ where $^T$ is the
transpose of a vector. For a vector of operators
$\mathbf{a}$, the  operator $\mathrm{col}(\mathbf{a},\mathbf{a}^\#)$ is denoted
$\breve{\mathbf{a}}$.   
For a complex matrix $A=(A_{ij})$,
$A^\#$, $A^T$, $A^\dagger$ denote, respectively, the matrix of complex
conjugates $(A_{ij}^*)$, the transpose matrix and the Hermitian adjoint
matrix. $[\cdot,\cdot]$ 
denotes the commutator of two operators in  $\mathfrak{H}$. $\tr[\cdot]$
denotes the trace of a matrix. $I$ is the identity matrix, and
$
J=
  \left[
  \begin{array}{rr}
    I & 0 \\ 0 & -I
  \end{array}
  \right].
$
 The quantum
expectation of an operator $V$ with respect to a state $\rho$, is
denoted $\langle V\rangle=\tr[\rho V]$~\cite{Parthasarathy-2012}.  
The cross-correlation of stationary quantum operator processes
$\mathbf{x}_j(t)$, $\mathbf{x}_k(t)$ will be denoted
$R_{\mathbf{x}_j,\mathbf{x}_k}(t)$, 
$
R_{\mathbf{x}_j,\mathbf{x}_k}(t)=\langle
  \mathbf{x_j}(t)\mathbf{x_k}^*(0)  \rangle.
$
The corresponding bilateral Laplace transform (understood in the sense of
tempered distributions when necessary) is denoted
$P_{\mathbf{x}_j,\mathbf{x}_k}(s)$,  
\begin{equation}
  \label{eq:27}
P_{\mathbf{x}_j,\mathbf{x}_k}(s)=\int_{-\infty}^{+\infty}e^{-s t}
R_{\mathbf{x}_j,\mathbf{x}_k}(t) dt.  
\end{equation}
For any two complex matrices $X_-$, $X_+$, we write 
$
  \Delta(X_-,X_+)\triangleq 
  \left[
    \begin{array}{cc} X_- & X_+ \\ X_+^\# & X_-^\# 
    \end{array}
  \right].
$
When $X_-$, $X_+$ are complex transfer functions $X_-(s)$, $X_+(s)$, the
corresponding stacking operation defines the transfer function 
$
  \Delta(X_-(s),X_+(s))\triangleq 
  \left[
    \begin{array}{cc} X_-(s) & X_+(s) \\ (X_+(s^*))^\# & (X_-(s^*))^\# 
    \end{array}
  \right].
$


\section{An open linear system model of a quantum communication
  channel}\label{sec:open-linear-passive} 

In the Heissenberg picture of quantum mechanics, an open quantum system can
be modeled as a linear system governed by an input 
field $\breve{\mathbf{b}}=\mathrm{col}(\mathbf{b},\mathbf{b}^\#)$
where $\mathbf{b}$ is a column vector of $n$ quantum noise processes,
$\mathbf{b}=\mathrm{col}(\mathbf{b}_1,\ldots,\mathbf{b}_n)$~\cite{HP-1984,ZJ-2013}. The
noise processes can be represented as annihilation operators on an appropriate
Fock space~\cite{HP-1984}, but 
from the system theory viewpoint they can be treated as quantum stochastic
processes. In this paper, it will be assumed that these input processes
represent Gaussian
white noise processes with zero mean, $\langle \mathbf{b}(t)\rangle =0$, and
the covariance 
\begin{eqnarray}
  \label{eq:8}
  \left\langle
  \left[
    \begin{array}{c}
     \mathbf{b}(t) \\ \mathbf{b}^\# (t) 
    \end{array}
  \right]  \left[
    \begin{array}{cc}
     \mathbf{b}^\dagger(t') \\ \mathbf{b}^T (t') 
    \end{array}
  \right]\right\rangle=
  \left[
    \begin{array}{cc}
I+\Sigma_{\mathrm{b}}^T & \Pi_{\mathrm{b}} \\
 \Pi_{\mathrm{b}}^\dagger & \Sigma_{\mathrm{b}}
\end{array}\right]\delta(t-t'),
\end{eqnarray}
where $\Sigma_{\mathrm{b}}$, $\Pi_{\mathrm{b}}$ are complex matrices with
the properties that $\Sigma_{\mathrm{b}}=\Sigma_{\mathrm{b}}^\dagger$,
$\Pi_{\mathrm{b}}^T=\Pi_{\mathrm{b}}$, and 
$\delta(t-t')$ is the $\delta$-function.

Using this notation, dynamics of an open
quantum system without scattering are described by a quantum stochastic
differential equation 
\begin{eqnarray}
  \label{dyn}
  \dot{\breve{\mathbf{a}}}&=&\breve A \breve{\mathbf{a}}+  \breve B
                              \breve{\mathbf{b}}, \quad \breve
                              {\mathbf{a}}(t_0)=\breve{\mathbf{a}}, \nonumber \\
  \breve{\mathbf{y}}&=&\breve C \breve{\mathbf{a}}+ \breve D \breve{\mathbf{b}}.
\end{eqnarray}
The column vector
$\breve{\mathbf{a}}=\mathrm{col}(\mathbf{a},\mathbf{a}^\#)$ is composed of
the column vector 
$\mathbf{a}=\mathrm{col}(\mathbf{a}_1, 
\ldots, \mathbf{a}_m)$ of annihilation  
operators on a certain Hilbert space $\mathfrak{H}$ and the column vector
$\mathbf{a}^\#=\mathrm{col}(\mathbf{a}_1^*, 
\ldots, \mathbf{a}_m^*)$ of the corresponding creation
operators on the same Hilbert space. Also,
$\breve{\mathbf{y}}=\mathrm{col}(\mathbf{y},\mathbf{y}^\#)$ denotes the
output field of the system that carries away information about the system
interacting with the input field $\breve{\mathbf{b}}$. The matrices
$\breve{A}$, $\breve{B}$, $\breve{C}$, $\breve{D}$ are 
partitioned accordingly, as 
\begin{eqnarray*}
  \label{eq:4}
\breve{A}&=&\Delta(A_-,A_+), \quad   
\breve{B}=\Delta(B_-,B_+), \\
\breve{C}&=&\Delta(C_-,C_+), \quad   
             \breve{D}=\Delta(D_-,D_+).
\end{eqnarray*}
A detailed discussion 
about open linear quantum systems can be found in
references~\cite{JG-2010,GJN-2010,JNP-2008,ZJ-2013}. 


In this paper, we are concerned with the situation where the system
(\ref{dyn}) models a quantum communication channel, and
$\breve{\mathbf{b}}$ and $\breve{\mathbf{y}}$ describe the input and output
signals of this channel. Furthermore, similarly to~\cite{UJ2a} we consider
a class of passive communication channels (\ref{dyn}) whose properties
make them analogous to classical passive systems~\cite{JG-2010}. In a
passive quantum system, 
$A_+=0$, $B_+=0$, $C_+=0$, and $D_+=0$. That is, the 
dynamics of $\mathbf{a}$ are governed by the input $\mathbf{b}$
consisting of annihilation operators only, and the
dynamics of $\mathbf{a}^\#$ are governed by the input $\mathbf{b}^\#$
consisting of creation operators only. Passivity reflects the fact that the
Hamiltonian of the system and its coupling with the environment only allow
dissipation of energy.   

With the above assumptions, the output field of the passive system
(\ref{dyn}) can be written as 
\begin{eqnarray}
\mathbf{y}(t) 
&=&C_- e^{A_-(t-t_0)}\mathbf{a}(t_0)
    +\int_{t_0}^tg(t-\tau)\mathbf{b}(\tau)d\tau,
    \nonumber \\ 
\mathbf{y}^\#(t) 
&=&C_-^\# e^{A_-^\#(t-t_0)}\mathbf{a}^\#(t_0)
    +\int_{t_0}^tg^\#(t-\tau)\mathbf{b}^\#(\tau)d\tau. \qquad
\label{conv}
\end{eqnarray}
Here we introduced the notation for the impulse response, associated with
the annihilation part of the system~\cite{ZJ-2013}, 
\begin{equation}
g(t)=
\begin{cases}
C_-e^{A_-t}B_-+\delta(t)I, & t\ge 0, \\
0, & t<0.  
\end{cases}
\label{ch.impresp}
\end{equation}
The transfer function of the passive system (\ref{dyn}) is then
\[
  \Gamma(s)=
  \left[
    \begin{array}{cc}
      G(s)& 0 \\
      0 & G(s^*)^\#
    \end{array}
  \right],
\]
where $G(s)=C_-(sI-A_-)^{-1}B_-+I$. Since $B_-=-C_-^\dagger$, the transfer
functions $G(s)$ and $\Gamma(s)$ are square matrices. 

Not every system of the form (\ref{dyn}) corresponds to physical quantum  
dynamics. For this to be true, the system must preserve the canonical
commutation relations during its evolution~\cite{SP-2012,JNP-2008}. This
property translates to a formal requirement~\cite{SP-2012,ZJ-2013} that for
a physically realizable system (\ref{dyn}) it must hold that
\begin{equation}
  \label{eq:1}
G(s)G(-s^*)^\dagger=I, \quad  \Gamma(s)J\Gamma(-s^*)^\dagger=J. 
\end{equation}

We will be concerned with stationary behaviours of the systems
under consideration. Suppose that the matrix $A_-$ is stable, then the
stationary component of the system output is obtained 
from (\ref{conv}) by letting $t_0\to -\infty$: 
\begin{eqnarray}
\mathbf{y}(t)&=& 
  \int_{-\infty}^{+\infty}g(t-\tau)\mathbf{b}(\tau)d\tau, \nonumber \\
\mathbf{y}^\#(t)&=& \int_{-\infty}^{+\infty}g^\#(t-\tau)\mathbf{b}^\#(\tau)d\tau.
\label{conv.1}
\end{eqnarray}
The upper limit of integration has been changed to $+\infty$ since $g(t)$
is causal by definition. Since the matrix $A_-$ is Hurwitz,
$P_{\mathbf{y}_j,\mathbf{y}_k}(s)$ is well defined on the imaginary axis and  
$P_{\mathbf{y}_j,\mathbf{y}_k}(s)|_{s=i\omega}=P_{\mathbf{y}_j,\mathbf{y}_k}(i\omega)$, 
where  the expression on the left-hand side refers to the
bilateral Laplace transform (\ref{eq:27}) and the expression on
right-hand side is the Fourier transform of 
$R_{\mathbf{y}_j,\mathbf{y}_k}(t)$. Both expressions are usually referred
to as the cross power spectrum density (cross PSD)~\cite{Kailath-1981}. 
It is easy to obtain that the power spectrum density matrix
of the output  $\mathbf{y}(t)$,  
$P_{\mathbf{y},\mathbf{y}}(s)=(P_{\mathbf{y}_j,\mathbf{y}_k}(s))_{j,k=1}^n$
is related to the power spectrum density matrix of the
noise $\breve{\mathbf{b}}$ in the standard manner~\cite{ZJ-2013}:
\begin{equation}
  P_{\mathbf{y},\mathbf{y}}(s)=
\Gamma(s)  \left[
    \begin{array}{cc}
I+\Sigma_{\mathrm{b}}^T & \Pi_{\mathrm{b}} \\
 \Pi_{\mathrm{b}}^\dagger & \Sigma_{\mathrm{b}}
\end{array}\right][\Gamma(-s^*)]^\dagger.  
  \label{PSD}  
\end{equation}

\section{Active equalization of passive quantum communication
  channels}\label{sec:equal-probl-annih}

In this section, we review the general equalization scheme introduced
in~\cite{UJ2a} and introduce the class of quantum systems which serve as
candidate coherent equalizers.

\begin{figure}[t]
\psfrag{Quantum}{}
\psfrag{channel}{$\Gamma (s)$}
\psfrag{equalizer}{$\Xi(s)$}
  \centering
  \psfrag{b}{$\breve b$}
  \psfrag{e}{$\breve e$}
  \psfrag{w}{$\breve w$}
  \psfrag{what}{$\breve y_w$}
  \psfrag{z}{$\breve z$}
  \psfrag{bhat}{$\breve {\hat b}$}
  \psfrag{zhat}{$\breve {\hat z}$}
  \psfrag{y}{$\breve y_b$}
  \includegraphics[width=0.9\columnwidth]{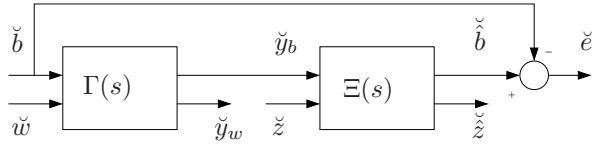}
  \caption{A general quantum communication system. The transfer function
    $\Gamma(s)$ represents the channel, and $\Xi(s)$ represents an
    equalizing filter.} 
  \label{fig:general}
\end{figure}

Consider the system in Fig.~\ref{fig:general} consisting of a
quantum channel $\Gamma(s)$ and a second quantum system acting as an equalizer.
The input field $b$ plays the role of a message signal 
transmitted through the channel, 
and $w$ denotes the vector comprised of quantum noises. It includes the
noise inputs that are necessarily 
present in the physically realizable system~\cite{JNP-2008,VP-2011}, as well as 
noises introduced by routing devices such as beam splitters. In terms of
the notation adopted in the previous section, we have
$\mathbf{b}=\mathrm{col}(b,w)$. This combined input and its adjoint signal are
transmitted through the quantum channel with the transfer function
$\Gamma(s)$, as described in the previous section, to produce the output 
$\breve{\mathbf{y}}=\mathrm{col}(y_b,y_w,y_b^\#,y_w^\#)$. The
input to the filter $\Xi(s)$ is comprised of the channel output field
components $\breve y_b=\mathrm{col}(y_b,y_b^\#)$. We emphasize that in contrast
to~\cite{UJ2a}, we consider filters that process both quadratures of the
channel output $\breve y_b$.     
 
Unlike the classical Wiener equalization problem, a coherent
filter must be realizable as a quantum system, and therefore it must preserve
canonical commutation relations. For this, the filter system must satisfy
the physical realizability conditions analogous to condition
(\ref{eq:1}). It is known~\cite{JNP-2008,VP-2011} that for this, 
additional noise inputs may need to be injected into the filter; the input
$z$ in Fig.~\ref{fig:general} symbolizes those additional noise inputs. As
in~\cite{UJ2a}, the added noise $z$ will be assumed to be in the Gaussian
vacuum state, i.e., the corresponding mean and covariance of $z$ are  
\begin{eqnarray}
\label{eq:12}
\langle z(t)\rangle =0, \quad 
  \left\langle
  \left[
    \begin{array}{c}
     z(t) \\ z^\# (t) 
    \end{array}
  \right]  \left[
    \begin{array}{cc}
     z^\dagger(t') \\ z^T (t') 
    \end{array}
   \right]\right\rangle=
  \left[
    \begin{array}{cc}
I & 0 \\
0 & 0
\end{array}\right]\delta(t-t'). \quad
\end{eqnarray}

With the additional noise $z$ injected into the filter, the filter system
can be regarded as a mapping $\breve u \to \breve{\hat 
u}$, where $u=\mathrm{col}(y_b,z)$, $\hat u=\mathrm{col}(\hat b,\hat
z)$. The filter transfer
function $\Xi(s)$  can be partitioned accordingly: 
\begin{eqnarray}
  \label{eq:3}
  \Xi(s)= \Delta( H(s), T(s)) 
= 
    \left[
    \begin{array}{cc}
     H(s) & T(s) \\
     T(s^*)^\# & H(s^*)^\#
    \end{array}
  \right].
\end{eqnarray}
The physical realizability condition for the filter is
analogous to  (\ref{eq:1}), 
\begin{equation}
  \label{eq:7}
  \Xi(s)J\Xi(-s^*)^\dagger = J.
\end{equation}
The set of physically realizable equalizers $\Xi(s)$, (i.e., filters satisfying condition (\ref{eq:7})) will be denoted
$\mathcal{H}$. In the sequel, we will also consider a subset of the set
$\mathcal{H}$ consisting of constant complex $J$-symplectic matrices
$\Xi=\Delta(H,T)$\footnote{A matrix $\Xi$ is $J$-symplectic if $\Xi
  J\Xi^\dagger =J$.}. The set of such matrices  will 
be denoted $\mathcal{H}_c$.   

Reference~\cite{UJ2a} proposed an approach to the design of coherent equalizers
which was analogous to the classical mean-square equalization scheme. It
aimed to compute a physically realizable transfer 
function of a filter by minimizing the power spectrum density
$P_{e,e}(i\omega)$ of the error operator $e(t)=\hat b(t)-b(t)$. In this
paper, the optimal equalization objective is defined in a similar manner,
as  
\begin{eqnarray}
  \label{eq:6}
  \label{eq:6'}
  \min_{\Xi\in \mathcal{H}}\sup_\omega \tr P_{e,e}(i\omega).
\end{eqnarray}
The outer optimization operation is to be carried out over the set
$\mathcal{H}$ of filters which satisfy~(\ref{eq:7}). Thus, (\ref{eq:6})
represents a constrained optimization problem.   

The problem in~\cite{UJ2a} can be regarded as a special case of the problem
(\ref{eq:6}) in which optimizers $\Xi(s)$ are constrained to 
those of the form $\Xi(s)=\Delta(H(s),0)$. Let $\mathcal{H}_p$ denote the
class of such filters,
\[
\mathcal{H}_p=\{\Xi\in\mathcal{H}\colon \Xi(s)=\Delta(H(s),0)~ (\exists H(s))\}.
\]
Note that for $\Xi\in\mathcal{H}_p$, the condition (\ref{eq:7}) reduces to
the condition $H_p(s)H_p(-s^*)^\dagger=I$. Clearly, we have
\begin{eqnarray}
  \label{eq:6.ap}
  \min_{\Xi\in \mathcal{H}}\sup_\omega\tr P_{e,e}(i\omega)\le \min_{\Xi\in
    \mathcal{H}_p}\sup_\omega \tr P_{e,e}(i\omega). 
\end{eqnarray}
In the next section we will consider a situation where the
optimal values of the two problems are equal. Also, we will discuss a
version of the coherent equalization problem where the inequality in 
(\ref{eq:6.ap}) is a strict inequality. 
 

Consider the partitions of the 
transfer functions $H(s)$ and $T(s)$ compatible with the 
partitions of the filter input and output operators $\mathrm{col}(y_b,z)$
$\mathrm{col}(\hat b,\hat z)$ and the corresponding partitions of the
adjoint operators:   
\begin{eqnarray}
  \label{eq:98a}
  H(s)=
  \left[
    \begin{array}{cc}
H_{11}(s) & H_{12}(s)\\H_{21}(s) & H_{22}(s)
    \end{array}
  \right], \quad
T(s)&=&\left[
  \begin{array}{cc}
T_{11}(s) & T_{12}(s)\\
T_{21}(s) & T_{22}(s)
  \end{array}\right]. \qquad 
\end{eqnarray}
Also, consider the partition of the 
transfer functions $G(s)$ compatible with the partition $\mathbf{b}=\mathrm{col}(b,w)$,
$\mathbf{y}=\mathrm{col}(y_b,y_w)$:
\begin{eqnarray}
  \label{eq:98}
  G(s)&=&
  \left[
    \begin{array}{cc}
G_{11}(s) & G_{12}(s)\\
G_{21}(s) & G_{22}(s)\\
    \end{array}
  \right].
\end{eqnarray}
The covariance matrix of the input $\breve{\mathbf{b}}$ is
assumed to be partitioned accordingly, as 
\begin{equation}
  \label{eq:25}
   \left[ \begin{array}{cc}
I+\Sigma_{\mathrm{b}}^T & \Pi_{\mathrm{b}} \\
 \Pi_{\mathrm{b}}^\dagger & \Sigma_{\mathrm{b}}
    \end{array}\right]=
\left[\begin{array}{cccc}
 I+\Sigma_b^T &0 & \Pi_b & 0 \\
 0 & I+\Sigma_w^T & 0 & 0 \\                       
  \Pi_b^\dagger & 0 & \Sigma_b & 0 \\
  0 & 0 & 0 & \Sigma_w
\end{array}\right].  
\end{equation}
The matrix on the right-hand side reflects a standard assumption
that the message signal $b$ and the noise signal $w$ are not
correlated. Also (\ref{eq:25}) reflects the standing assumption in
this paper that $\langle w(t)w^T(t')\rangle =\langle w(t)^\# w^\dagger
(t')\rangle =0$. This assumption about the noise field $w$ is 
the same as the corresponding assumption made 
in~\cite{UJ2a}. However, in contrast to~\cite{UJ2a} here we do not generally assume that $\Pi_b=0$.

With these assumptions, we obtain using (\ref{PSD}) that
\begin{eqnarray}
\label{eq:121}
\lefteqn{P_{e,e}(s,\Xi(s))} && \nonumber \\
&=&
(H_{11}(s)G_{11}(s)-I)(I+\Sigma_b^T)(G_{11}(-s^*)^\dagger 
H_{11}(-s^*)^\dagger -I) 
\nonumber \\
&+&
H_{11}(s)G_{12}(s)(I+\Sigma_w^T)G_{12}(-s^*)^\dagger H_{11}(-s^*)^\dagger \nonumber \\
&+&
    H_{12}(s)H_{12}(-s^*)^\dagger \nonumber \\
&+& T_{11}(s)G_{11}(s^*)^\#\Sigma_bG_{11}(-s)^TT_{11}(-s^*)^\dagger \nonumber\\
&+& T_{11}(s)G_{12}(s^*)^\#\Sigma_wG_{12}(-s)^TT_{11}(-s^*)^\dagger \nonumber\\
&+& T_{11}(s)G_{11}(s^*)^\#\Pi_b^\dagger(G_{11}(-s^*)^\dagger H_{11}(-s^*)^\dagger -I) \nonumber\\
&+& (H_{11}(s)G_{11}(s)-I)\Pi_bG_{11}(-s)^TT_{11}(-s^*)^\dagger.
\end{eqnarray}
Here, we used the notation $P_{e,e}(s,\Xi(s))$ to specify the transfer
function $\Xi(s)$ of the system used as the filter in the system in
Fig.~\ref{fig:general}.  
Also, the constraint (\ref{eq:7}) can be expanded as follows, 
\begin{eqnarray}
  \label{eq:9}
&&
H_{11}(s)H_{11}(-s^*)^\dagger+H_{12}(s)H_{12}(-s^*)^\dagger \nonumber \\
  &&\hspace{2ex} - T_{11}(s)T_{11}(-s^*)^\dagger-T_{12}(s)T_{12}(-s^*)^\dagger=I, \\
&&
H_{11}(s)H_{21}(-s^*)^\dagger+H_{12}(s)H_{22}(-s^*)^\dagger \nonumber \\
  &&\hspace{2ex} - T_{11}(s)T_{21}(-s^*)^\dagger-T_{12}(s)T_{22}(-s^*)^\dagger=0, \label{eq:10}
\\
&& 
H_{21}(s)H_{21}(-s^*)^\dagger+H_{22}(s)H_{22}(-s^*)^\dagger \nonumber \\
  &&\hspace{2ex} -
   T_{21}(s)T_{21}(-s^*)^\dagger-T_{22}(s)T_{22}(-s^*)^\dagger=I. \label{eq:11}
  \\
&& 
H_{11}(s)T_{11}(-s)^T+H_{12}(s)T_{12}(-s)^T \nonumber \\
  &&\hspace{2ex} - T_{11}(s)H_{11}(-s)^T-T_{12}(s)H_{12}(-s)^T=0, \label{eq:5}
\\
&& 
H_{11}(s)T_{21}(-s)^T+H_{12}(s)T_{22}(-s)^T \nonumber \\
  &&\hspace{2ex} - T_{11}(s)H_{21}(-s)^T-T_{12}(s)H_{22}(-s)^T=0, \label{eq:18}
\\
&& 
H_{21}(s)T_{21}(-s)^T+H_{22}(s)T_{22}(-s)^T \nonumber \\
  &&\hspace{2ex} - T_{21}(s)H_{21}(-s)^T-T_{22}(s)H_{22}(-s)^T=0, \label{eq:24}
\end{eqnarray}

From (\ref{eq:121}), we observe that the spectral density function $
P_{e,e}(s,\Xi(s))$ depends only on the
transfer functions $H_{11}(s)$, $H_{12}(s)$ and $T_{11}(s)$ within
$\Xi(s)$. Therefore, similarly to~\cite{UJ2a} a
two-step procedure can be employed to solve the constrained optimization problem
(\ref{eq:6}). The first step of this procedure is to minimize the power
spectrum density objective $\sup_\omega P_{e,e}(i\omega,\Xi(i\omega))$
subject to the following relaxed version of (\ref{eq:9}) as the optimization constraint,
\begin{eqnarray} 
  \label{eq:28.a}
&& H_{11}(i\omega)H_{11}(i\omega)^\dagger+H_{12}(i\omega)H_{12}(i\omega)^\dagger
\nonumber \\ 
                   && \hspace{4ex}- T_{11}(i\omega)T_{11}(i\omega)^\dagger - T_{12}(i\omega)T_{12}(i\omega)^\dagger = I. 
\end{eqnarray}
Indeed, if an optimal equalizer in the problem (\ref{eq:6}) 
exists, it must necessarily satisfy (\ref{eq:28.a}). Let $\mathcal{H}_\omega$ denote the set of transfer functions of the form
$\Xi(s)=\Delta(H(s),T(s))$ which satisfy the condition (\ref{eq:28.a}) for a
given $\omega$. Also, let 
$\bar{\mathcal{H}}=\cap_{\omega}\mathcal{H}_\omega$, i.e, 
$\bar{\mathcal{H}}$ is a set of transfer functions of the form
$\Xi(s)=\Delta(H(s),T(s))$ which satisfy the condition (\ref{eq:28.a}) for
every $\omega$. Thus, $\mathcal{H}\subseteq \bar{\mathcal{H}}$, and we obtain
the following lower bound on (\ref{eq:6}), 
\begin{eqnarray}
  \label{eq:6.relaxed}
  \min_{\Xi\in \bar{\mathcal{H}}}\sup_\omega \tr P_{e,e}(i\omega)
\le   \min_{\Xi\in \mathcal{H}}\sup_\omega\tr P_{e,e}(i\omega).
\end{eqnarray}
In the second step, the set of solutions of the problem on the left-hand
side of  (\ref{eq:6.relaxed}) must be reduced to select only those
$\Xi(s)$ which satisfy all of the constraints (\ref{eq:9})-(\ref{eq:24}). If such
$\Xi(s)$ can be selected, then the lower bound (\ref{eq:6.relaxed}) is tight. 
In the next section, we will use this procedure to investigate whether
expanding the class of filters from passive filters of the form
$\Delta(H(s),0)$ (as considered in \cite{UJ2a}) to filters of the form
$\Delta(H(s),T(s))$ leads to an improved mean-square error.

\section{The main results}\label{sec:main-results}

This section presents the main results of the paper. In
Section~\ref{sec:equal-quant-chann}, an equalization
problem is presented in which expanding the set of filters from
$\mathcal{H}_p$ to $\mathcal{H}$ does not reduce the optimal power spectrum
density guaranteed by passive filters. In this problem  $\langle
b(t)b^T(t')\rangle =0$, i.e., $\Pi_b=0$.
Next, in Section~\ref{sec:equal-chann-with} the
coherent equalization problem for static channels will
be analyzed in which the input filed $b$ is squeezed, i.e.,  $\langle
b(t)b^T(t')\rangle =\Pi_b\delta(t-t')$, $\Pi_b\neq 0$. It will be shown
that any optimal equalizer arising in this problem utilizes both inputs
$y_b$ and $y_b^\#$. 

To present these results, we restrict the class of systems under
consideration to systems with scalar operator inputs. Accordingly, the transfer functions $G_{ij}$, $H_{ij}$, $T_{ij}$
are assumed to be scalar. The constraint (\ref{eq:28.a}) reduces to
\begin{eqnarray}
  \label{eq:28}
  |H_{11}(i\omega)|^2+|H_{12}(i\omega)|^2 
                         - |T_{11}(i\omega)|^2- |T_{12}(i\omega)|^2  = 1. \quad
\end{eqnarray}
Also, $\Sigma_b$, $\Sigma_w$, $\Pi_b$, etc., are scalars. To emphasize the
latter fact, we will use the 
lower case notation, i.e., $\Sigma_b=\sigma_b^2$, $\Sigma_w=\sigma_w^2$,
$\Pi_b=\pi_b$, where $\sigma_b$, $\sigma_w$, $\pi_b$ are constant, and
$\sigma_b$, $\sigma_w$ are real. In this
case, $P_{e,e}(i\omega, \Xi(i\omega))$ is scalar,
\begin{eqnarray}
\label{eq:121.a}
\lefteqn{P_{e,e}(i\omega, \Xi(i\omega)) =
(1+\sigma_b^2)|H_{11}(i\omega)G_{11}(i\omega)-1|^2} && 
\nonumber \\
&& +(1+\sigma_w^2)
|H_{11}(i\omega)|^2|G_{12}(i\omega)|^2 +|H_{12}(i\omega)|^2 \nonumber \\
&& +
(\sigma_b^2
   |G_{11}(-i\omega)|^2+\sigma_w^2|G_{12}(-i\omega)|^2)|T_{11}(i\omega)|^2.\qquad
  \nonumber \\
&& +2\mathrm{Re}\left[\pi_b(H_{11}(i\omega)G_{11}(i\omega)-1)G_{11}(-i\omega)T_{11}(i\omega)^*\right].  \qquad
\end{eqnarray}

\subsection{Equalization of scalar passive quantum channels when
  $\pi_b=0$}\label{sec:equal-quant-chann}  

When an equalizing filter is restricted to have
$T_{11}(s)=0$, $T_{12}(s)=0$, the function (\ref{eq:121.a})
reduces to the function used as an optimization objective in~\cite{UJ2a}.
Indeed, letting $T_{11}(s)=0$, $T_{12}(s)=0$ means that the output channel
$\hat b$ of the filter is 
$
\hat b=H_{11}(s) b+ H_{12}(s) z. 
$
The following results confirm that in this case optimization of the error power
spectrum density can be reduced to optimization over the set
$\mathcal{H}_p$.  

\begin{lemma}
  \label{L2}
If $T_{11}(s)=0$, $T_{12}(s)=0$ in the partition (\ref{eq:98a}) of
$\Xi_a(s)$ which satisfies condition (\ref{eq:9}), then, there exists
$\Xi_p(s)\in \mathcal{H}_p$ such that 
\begin{equation}
  \label{eq:41}
  P_{e,e}(i\omega,\Xi_a(i\omega))=P_{e,e}(i\omega,\Xi_p(i\omega))
  \quad \forall \omega\in \mathbf{R}^1.
\end{equation}
\end{lemma}


\begin{remark}
  The transfer function $\Xi_p$ 
  may not be
  causal. \hfill$\Box$

\end{remark}

We now discuss a method for obtaining a passive physically realizable
filter which attains an optimal value in the problem (\ref{eq:6}).

Since $\pi_b=0$, the scalar function $P_{e,e}(i\omega, \Xi(i\omega))$ becomes 
\begin{eqnarray}
\label{eq:121'}
\lefteqn{P_{e,e}(i\omega, \Xi(i\omega)) =
(1+\sigma_b^2)|H_{11}(i\omega)G_{11}(i\omega)-1|^2} && 
\nonumber \\
&& +(1+\sigma_w^2)
|H_{11}(i\omega)|^2|G_{12}(i\omega)|^2 +|H_{12}(i\omega)|^2 \nonumber \\
&& +
(\sigma_b^2 |G_{11}(-i\omega)|^2+\sigma_w^2|G_{12}(-i\omega)|^2)|T_{11}(i\omega)|^2.\qquad 
\end{eqnarray}

First we establish a result about an auxiliary point-wise optimization problem
\begin{equation}
  \label{eq:37}
 V_\omega \triangleq \min_{\Xi\in \mathcal{H}_{c,0}}  P_{e,e}(i\omega, \Xi).
\end{equation}
Here $\Xi=\Delta(H,T)$ is a complex constant 
matrix composed of complex matrices $H$, $T$, partitioned in the same way
as in (\ref{eq:98a}), and the notation 
$\mathcal{H}_{c,0}$ refers to the set of constant matrices
$\Xi=\Delta(H,T)$ which satisfy the condition
\begin{eqnarray}
  \label{eq:28.om.const.0}
  \label{eq:28.om.const}
  |H_{11}|^2+|H_{12}|^2 - |T_{11}|^2 - |T_{12}|^2 = 1.
\end{eqnarray}
Furthermore, $P_{e,e}(i\omega, \Xi)$ refers to the value on the
right-hand side of (\ref{eq:121'}) in which the components of
$\Xi(i\omega)$ are replaced with the corresponding components of 
$\Xi$.

Without loss of generality, it will be assumed that
\begin{eqnarray}
 && \psi(i\omega)\triangleq
  \sigma_b^2|G_{11}(i\omega)|^2+\sigma_w^2|G_{12}(i\omega)|^2>0, \nonumber \\
  && |G_{11}(i\omega)|>0 \quad \forall \omega\in\mathbf{R}^1. 
    \label{eq:39}
\end{eqnarray}

\begin{lemma}\label{L1}
  Suppose $\pi_b=0$. Let $\omega$ be fixed and
let a matrix $\Xi_\omega=\Delta(H_\omega,T_\omega)$ attain the 
minimum in the problem (\ref{eq:37}). Then $T_{\omega,11}=0$.  
Furthermore, the following statements hold. 
\begin{enumerate}[1)]
\item 
If 
\begin{equation}
  \label{eq:22}
  \frac{(1+\sigma_b^2)|G_{11}(i\omega)|}{1+\psi(i\omega)}> 1,
\end{equation}
then $H_{\omega,12}=0$,
$H_{\omega,11}=\frac{(1+\sigma_b^2)G_{11}(i\omega)^*}{1+\psi(i\omega)}$,
$|T_{\omega,12}|^2=\frac{(1+\sigma_b^2)^2|G_{11}(i\omega)|^2}{(1+\psi(i\omega))^2}-1>0$, 
 and
 \begin{equation}
   \label{eq:46}
V_\omega =
(1+\sigma_b^2)\left(1-\frac{(1+\sigma_b^2)|G_{11}(i\omega)|^2}{1+\psi(i\omega)}\right).   
 \end{equation}

\item
If 
\begin{equation}
\label{eq:38}
  \frac{\psi(i\omega)}{1+\psi(i\omega)}<\frac{(1+\sigma_b^2)|G_{11}(i\omega)|}{1+\psi(i\omega)}\le
  1, 
\end{equation}
then $H_{\omega,11}=\frac{G_{11}(i\omega)^*}{|G_{11}(i\omega)|}$,
$H_{\omega,12}=0$, $T_{\omega,12}=0$. Furthermore,
\begin{eqnarray}
  \label{eq:51}
 V_\omega =
  (1+\sigma_b^2)+1+\psi(i\omega)-2(1+\sigma_b^2)|G_{11}(i\omega)|. \quad
\end{eqnarray}

\item
If 
\begin{equation}
\label{eq:26} 
(1+\sigma_b^2)|G_{11}(i\omega)|\le \psi(i\omega),
\end{equation}
then $H_{\omega,11}=\frac{(1+\sigma_b^2)G_{11}(i\omega)^*}{\psi(i\omega)}$,
$T_{\omega,12}=0$ and $H_{\omega,12}$ is such that
$|H_{\omega,12}|^2=1-\frac{(1+\sigma_b^2)^2|G_{11}(i\omega)|^2}{\psi(i\omega)^2}$. Furthermore,
\begin{equation}
  \label{eq:52}
V_\omega =
(2+\sigma_b^2)-\frac{(1+\sigma_b^2)^2|G_{11}(i\omega)|^2}{\psi(i\omega)}.  
\end{equation}
\end{enumerate}
\end{lemma}

The proof of Lemma~\ref{L1} is based on the method of Lagrange multiplier and
is omitted for brevity.  

We now
establish a connection between the point-wise optimization problem
(\ref{eq:37}) and the underlying problem (\ref{eq:6}) in the case where the channel $G(s)$ satisfies the condition
\begin{equation}
\label{eq:38.a}
  \frac{(1+\sigma_b^2)|G_{11}(i\omega)|}{1+\psi(i\omega)}\le 1, \quad
  \forall \omega\in\mathbf{R}^1.
\end{equation}
According to Lemma~\ref{L1}, under this condition,
$T_{\omega,11}=T_{\omega,12}=0$ in any optimal point $\Xi_\omega$ of the
problem (\ref{eq:37}).

Note that since $G_{11}(s)$ and
$G_{12}(s)$ are rational transfer functions, the 
expressions for 
$H_{\omega,11}$, $H_{\omega,12}$ obtained in Lemma~\ref{L1} are also
rational functions of $\omega$. Therefore using standard techniques, one
can obtain transfer functions $H_{11}(s)$, $H_{12}(s)$ which match the 
frequency responses $H_{\omega,11}$, $H_{\omega,12}$ obtained in
Lemma~\ref{L1}. Secondly, under condition (\ref{eq:38.a}),
$T_{\omega,11}=T_{\omega,12}=0$, and we will show next that the only 
rational transfer functions $T_{11}(s)$, $T_{12}(s)$ that satisfy this
requirement are $T_{11}(s)=0$, $T_{12}(s)=0$. The
remaining entries of the matrix $\Xi_\omega=\Delta(H_\omega,T_\omega)$ do
not affect the optimal value of the objective function $P(i\omega,\Xi_\omega)$  
They can be selected so that a point-wise optimal
solution $\Xi_\omega$ of the problem (\ref{eq:37}) represents a frequency
response of a rational transfer function. 

\begin{theorem}
  \label{T2}
  Suppose $\pi_b=0$ and condition (\ref{eq:38.a}) is satisfied. If
  $\Xi_\omega=\Delta(H_\omega,T_\omega)$ attains the minimum in
  (\ref{eq:37}) for every $\omega\in\mathbf{R}^1$ and there exists
  a rational transfer function $\bar\Xi(s)=\Delta(\bar H(s), \bar
  T(s))$ such that $\bar\Xi(i\omega)=\Xi_\omega$ $\forall\omega$, then:
  \begin{enumerate}[(i)]
  \item
    $\bar T_{11}(s)=0$, $\bar T_{12}(s)=0$.
  \item
    $\bar\Xi(s)\in \bar{\mathcal{H}}$ and attains minimum in the problem
\begin{equation}
  \label{eq:37.sup}
  \min_{\Xi\in \bar{\mathcal{H}}}\sup_\omega
 P_{e,e}(i\omega,\Xi(i\omega)),  
 \end{equation}
  \item
    In addition, if $\bar H_{11}(s)$, $\bar H_{12}(s)$ satisfy
\begin{eqnarray}
&&
   \bar H_{11}(s)\bar H_{11}(-s^*)^\dagger+\bar H_{12}(s)\bar
   H_{12}(-s^*)^\dagger =1, \quad
   \label{eq:9p}
\end{eqnarray}
    then an optimal filter in the problem (\ref{eq:6}) can
    be found within the class of passive filters $\mathcal{H}_p$.
  \end{enumerate}
\end{theorem}

When either condition~(\ref{eq:38}) or condition~(\ref{eq:26}) hold for all
$\omega$, Theorem~\ref{T2} provides a constructive method for deriving a
passive (possibly noncausal) optimal transfer function which solves the
underlying 
optimization problem (\ref{eq:6}). Under either of these conditions 
Lemma~\ref{L1} yields single closed form expressions for $H_{\omega,11}$ and
$H_{\omega,12}$. This allows to obtain transfer functions
$H_{11}(s)$, $H_{12}(s)$ such that $H_{11}(i\omega)=H_{\omega,11}$,
$H_{12}(i\omega)=H_{\omega,12}$ using standard factorization
techniques. Next, using the obtained  
  $H_{11}(s)$, $H_{12}(s)$, a transfer
  function $H_p(s)$ can be constructed as described in Lemma~\ref{L2}. The
  resulting transfer function $\Xi_p(s)=\Delta(H_p(s),0)$ is 
  physically realizable, and
  $\Xi_\omega=\Xi_p(i\omega)=\Delta(H_p(i\omega),0)$ satisfies
  all conditions of Theorem~\ref{T2}. Therefore $\Xi_p(s)=\Delta(H_p(s),0)$
  is an optimal passive equalizer for the problem (\ref{eq:6}).

\subsection{Equalization of scalar static passive quantum channels when
  $\pi_b\neq 0$: An optimal filter is an active quantum
  system}\label{sec:equal-chann-with} 

In this section, the optimization problem (\ref{eq:6}) is revisited for a
squeezed noise input $b$, i.e., when $\pi_b\neq 0$. To demonstrate
that in this case active filters may provide an advantage over passive
filters, it will suffice to consider a static 
quantum channel. That is, in this section we assume that the transfer
function $G(s)$ is a constant unitary matrix:
\begin{eqnarray}
  \label{eq:1.bs}
  \left[
    \begin{array}{c}
      y_b \\ y_w
    \end{array}
  \right]=  G \left[
    \begin{array}{c}
      b \\  w
    \end{array}
  \right], \quad G=\left[
    \begin{array}{cc}G_{11} & G_{12} \\
         -e^{i\theta}G_{12}^* & e^{i\theta}G_{11}^*
    \end{array}
  \right]; \quad
\end{eqnarray}
where $\theta\in[0,2\pi]$ and $|G_{11}|^2+|G_{12}|^2=1$.

The quantity $\psi=\sigma_b^2|G_{11}|^2+\sigma_w^2|G_{12}|^2$ 
is constant in this case. Also, as in the previous section, assume that
$|G_{11}|>0$. Since the covariance matrix $\left[
  \begin{array}{cc}
    1+\sigma_b^2 & \pi_b \\ \pi_b^* & \sigma_b^2
  \end{array}
\right]$ is positive definite, and $\pi_b\neq 0$, then it must hold that
$\sigma_b^2>0$. Together with the assumption that $|G_{11}|>0$ this implies
$\psi>0$. That is, conditions (\ref{eq:39}) are satisfied in this section
as well. In addition, we will assume in this section that $|G_{11}|^2<1$;
this implies that $|G_{12}|^2>0$ since $|G_{11}|^2+|G_{12}|^2=1$. These
assumptions mean that we do not consider unrealistic situations where the
channel is noiseless or blocks transmission of the field $b$. 

Since all coefficients in (\ref{eq:1.bs}) are constants, in this section we
will suppress the variable $i\omega$ and write $P_{e,e}(\Xi)$ or
$P_{e,e}(\Xi(i\omega))$ for a $\Xi(s)=\Delta(H,T)$.

The main result of this section is as follows.

\begin{theorem}\label{Prop1}
  Suppose $0<|G_{11}|<1$ and $\pi_b\neq 0$.  
If a proper rational transfer function $\Xi_0=\Delta(H_0,T_0)\in \mathcal{H}$ is
an optimal filter in the problem (\ref{eq:6}) for a static channel
(\ref{eq:1.bs}), then it must hold that $T_{0,11}(s)\neq 0$. Furthermore,
the same optimal performance can be achieved using a static coherent filter
in which $T_{11}\neq 0$ and at least one of the coefficients $H_{12}$,
$T_{12}$ is equal to 0.   
\end{theorem}

From Theorem~\ref{Prop1}, it follows that a mean-square optimal estimate of
a scalar squeezed input $b$ transmitted via a static quantum channel can be
obtained using one of the following expressions 
\begin{eqnarray*}
\hat b=H_{11}b+T_{11}b^*+T_{12}z^*
\end{eqnarray*}
or
\begin{eqnarray*}
\hat b=H_{11}b+H_{12}z+T_{11}b^* .
\end{eqnarray*}
The coefficients of these filters are constant. They can be obtained from the
 auxiliary optimization problem 
\begin{equation}
  \label{eq:37.const}
 V \triangleq \min_{\mathcal{H}_{c,0}} P_{e,e}(\Xi)
\end{equation}
which can be solved directly using
the Lagrange multiplier technique
. 
Since for a static channel $G$ the cost of the equalization problem does
not depend on the frequency variable $\omega$ explicitly, the auxiliary
problem (\ref{eq:37.const}) is not parameterized by $\omega$. 
The optimization in (\ref{eq:37.const}) is carried out over the set of
constant matrices $\Xi=\Delta(H,T)$ subject to the constraint
(\ref{eq:28.om.const}).

\section{Conclusions}\label{Conclusions}
The paper has presented new results on the quantum counterpart of the
classical Wiener filtering approach to equalization of quantum
communication systems
introduced in~\cite{UJ2a}. It has focused on the question as to whether
the mean-square performance achievable by passive annihilation-only
filters can be improved by driving the filter's annihilation (respectively,
creation) dynamics by both annihilation and creation components of
the channel output. We have shown that in general, the answer to this
question depends on the characteristics of the channel input field.

When
the channel input field is in thermal state, i.e., $\langle
b(t)b^T(t')\rangle=0$, the paper has provided general conditions under which a
physically realizable coherent mean-square optimal equalizing filter can be
found within the class of passive 
(possibly noncausal) filters. In this case, optimal
equalization of the channel distortion and noise may be accomplished passively,
by dissipating  
energy in the filter output field. On the other hand, when the
scalar input field to a static channel is in a squeezed state so that $\langle
b(t)b(t')\rangle=\pi_b\delta(t-t')$, with $\pi_b\neq 0$, the
mean-square optimal estimate of the input field can only be obtained using
an active filter.

The requirement for physical realizability of a filter introduces a
critical constraint into the proposed optimization approach to 
quantum equalization. Due to this requirement, our result ascertaining the
possibility of using passive equalizers for thermal input fields is
limited in that the optimal passive filter constructed in
Section~\ref{sec:equal-quant-chann} may not be causal. One of the possible
directions for future work will be to address the requirement
for causality of synthesized equalizers in a systematic manner.

\bibliography{Val,irpnew}
\bibliographystyle{plain}

\end{document}